\documentclass[twocolumn,aps,showpacs,prl,preprintnumbers,bibnotes10pt,superscriptaddress]{revtex4}
\usepackage{graphicx}
\usepackage{epsfig}

\usepackage[normalem]{ulem}

\begin{document}

\title{Sub-Doppler laser cooling of potassium atoms}
\author{M. Landini}
\email{landini@science.unitn.it} \affiliation{LENS and Dipartimento di Fisica e
Astronomia, Universit\'a di Firenze, 50019 Sesto Fiorentino, Italy} \affiliation{INFN,
Sezione di Firenze, 50019 Sesto Fiorentino, Italy} \affiliation{INO-CNR, 50019 Sesto
Fiorentino, Italy} \affiliation{Dipartimento di Fisica, Universit\`{a} di Trento, 38123
Povo, Trento, Italy}
\author{S. Roy}
\affiliation{LENS and Dipartimento di Fisica e Astronomia, Universit\'a di Firenze, 50019
Sesto Fiorentino, Italy}
\author{L. Carcagn\'{i}}
\affiliation{LENS and Dipartimento di Fisica e Astronomia, Universit\'a di Firenze, 50019
Sesto Fiorentino, Italy} \affiliation{INO-CNR, 50019 Sesto Fiorentino, Italy}
\author{D. Trypogeorgos}
\altaffiliation{Present address: Clarendon Laboratory, University of Oxford, Parks Road,
Oxford, OX1 3PU, UK} \affiliation{LENS and Dipartimento di Fisica e Astronomia,
Universit\'a di Firenze, 50019 Sesto Fiorentino, Italy}
\author{M. Fattori}
\affiliation{LENS and Dipartimento di Fisica e Astronomia, Universit\'a di Firenze, 50019
Sesto Fiorentino, Italy} \affiliation{INFN, Sezione di Firenze, 50019 Sesto Fiorentino,
Italy} \affiliation{INO-CNR, 50019 Sesto Fiorentino, Italy}
\author{M. Inguscio}
\affiliation{LENS and Dipartimento di Fisica e Astronomia, Universit\'a di Firenze, 50019
Sesto Fiorentino, Italy} \affiliation{INFN, Sezione di Firenze, 50019 Sesto Fiorentino,
Italy} \affiliation{INO-CNR, 50019 Sesto Fiorentino, Italy}
\author{G. Modugno}
\affiliation{LENS and Dipartimento di Fisica e Astronomia, Universit\'a di Firenze, 50019
Sesto Fiorentino, Italy} \affiliation{INFN, Sezione di Firenze, 50019 Sesto Fiorentino,
Italy} \affiliation{INO-CNR, 50019 Sesto Fiorentino, Italy}

\begin{abstract}
We investigate sub-Doppler laser cooling of bosonic potassium isotopes, whose small
hyperfine splitting has so far prevented cooling below the Doppler temperature. We find
instead that the combination of a dark optical molasses scheme that naturally arises in
this kind of systems and an adiabatic ramping of the laser parameters allows to reach
sub-Doppler temperatures for small laser detunings. We demonstrate temperatures as low as
(25$\pm$3)$\mu$K and (47$\pm$5)$\mu$K in high-density samples of the two isotopes
$^{39}$K and $^{41}$K, respectively. Our findings will find application to other atomic
systems.
\end{abstract}

\pacs{37.10.De; 37.10.Vz}

\date{\today}
\maketitle

Sub-Doppler laser cooling of neutral atoms~\cite{nobel} is a key technique for the
production of ultracold and quantum gases. It allows for atoms to be cooled to
temperatures below the Doppler limit~\cite{hansch}, not far from the single photon recoil
energy. This favors the application of further cooling techniques, such as Raman or
evaporative cooling, to reach quantum degeneracy. It also realizes a fast and effective
cooling method for some classes of atomic interferometers and clocks \cite{metrologia}.
The sub-Doppler cooling mechanism arises whenever the atomic ground state has an internal
structure with state-dependent light shifts. Such a situation is typically accompanied by
a hyperfine structure of the excited state \cite{RMP}. The sub-Doppler cooling is
efficient only if the excited state has a hyperfine splitting $\Delta$ either much larger
than the natural linewidth $\Gamma$, like that for the alkalis Na, Rb and Cs, or smaller
than $\Gamma$, as for example in Sr \cite{strontium}. In the intermediate case of
$\Delta\sim\Gamma$, it can instead be hindered by the presence of heating forces or by
photon reabsorption \cite{foot}. The bosonic potassium isotopes fall into this latter
category \cite{bambini}, and no efficient sub-Doppler cooling has been observed so far
\cite{walker,fort,prevedelli,minardi39,sympathetic,inouye}.

We now instead find that sub-Doppler cooling can take place also in atoms like K, by
employing a near-detuned optical molasses and an appropriate strategy to tune the cooling
laser parameters. We observe that the natural depumping towards dark states taking place
in this kind of systems allows us to reach low temperatures even in high-density samples.
In experiments on the isotopes $^{39}$K and $^{41}$K, we achieve temperatures
substantially lower than those previously achieved, with an efficiency similar to that of
most other alkali species.

\begin{figure}[ht]
\begin{center}
\includegraphics[width=0.9\columnwidth] {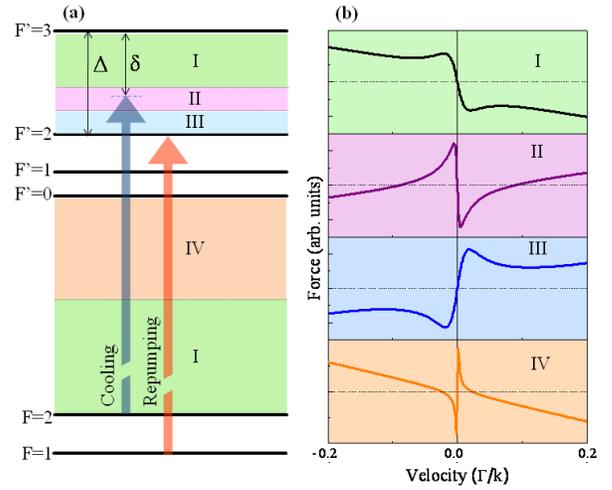}
\end{center}
\caption{Working regions for sub-Doppler cooling of bosonic Potassium. a) Level scheme
including the relevant hyperfine splitting $\Delta$ and the cooling laser detuning
$\delta$. b) Calculated cooling forces vs the atomic velocity in the various regions of
(a). Doppler cooling only takes place in regions I and IV, while sub-Doppler cooling is
active only in regions I and II.} \label{zones}
\end{figure}

In a two-level system, Doppler cooling arises when a laser is tuned below the atomic
transition frequency, where the atoms experience a friction force: $F=-\alpha v$. If the
presence of Zeeman sub-levels is taken into account~\cite{josab}, a much larger friction
arises for small velocities, leading to temperatures much lower than the Doppler limit
$k_B T_{D}=\hbar \Gamma/2$. While in principle, the lowest achievable sub-Doppler
temperatures are independent of the laser detuning $\delta$ \cite{proc}, the experiments
with large density samples are performed at large detunings, $\delta\gg\Gamma$. This
requirement arises from the need of keeping the scattering rate of photons by individual
atoms low, in such a way that spontaneously emitted photons may not disturb the cooling
process \cite{foot}. Most atomic systems cannot be modeled as simple two-level ones since
they feature a hyperfine structure like the one in Fig.~\ref{zones} that is relevant for
instance to Na, K and Rb. In this case, it is commonly thought that $\delta$ must also be
smaller than the main hyperfine splitting $\Delta$, since otherwise the presence of the
other excited states would turn the sub-Doppler mechanism into a heating one. As a matter
of fact, in the case of the bosonic K isotopes, where $\Delta\approx2\Gamma-3 \Gamma$, a
clear sub-Doppler cooling has not been experimentally observed. This can be understood
from the nature of the optical forces we have calculated for the level structure in
Fig.\ref{zones}: sub-Doppler cooling is active either very close to resonance, where
heating from photon reabsorption might be large, or for $\delta\gg\Delta$, where however
the velocity capture range is very low. Note that in the case of $\Gamma>\Delta$, such as
for $^{87}$Sr, the sub-Doppler cooling stays efficient also for $\delta>\Delta$
\cite{strontium}.

\begin{figure}[ht]
\begin{center}
\includegraphics[width=0.9\columnwidth] {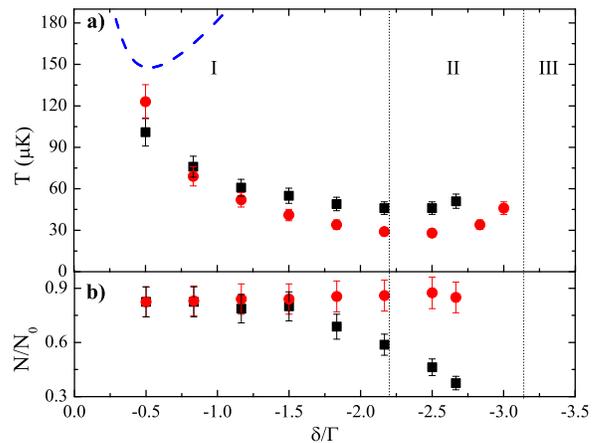}
\end{center}
\caption{Optimal sub-Doppler temperatures for $^{39}$K. a) Measured temperature without
(black squares) and with (red dots) the ramping strategy compared to the Doppler theory
(dashed line). The dotted lines separate the various regions as in Fig.~\ref{zones}. b)
Fraction of atoms remaining in the colder component without (black squares) and with (red
dots) the ramping strategy.} \label{K39}
\end{figure}

We have now realized that the presence of neighboring excited states, however, has also a
beneficial effect. Indeed it causes a natural depumping of the atomic population into a
dark state, such as the $F=1$ ground state as shown in Fig. \ref{zones}. The atoms can of
course be moved back into the bright $F=2$ state by the repumper laser but, differently
from a pure two-level system, this can be done in a controlled way. It is then possible
to adjust the fraction of atoms in the state coupled to the cooling laser in order to
optimize the cooling power, while keeping the reabsorption of spontaneously emitted
photons under control. Note that the possibility of controlling the population of the
bright state is absent if $\Delta\gg\Gamma$ unless an appropriate depumping laser is used
\cite{depumping}. This mechanism, which is widely used to trap atomic samples at
high-density in magneto-optical traps \cite{dark spot}, turns out to be the first
essential ingredient for sub-Doppler cooling when $\Delta\approx\Gamma$, since it allows
to reach low temperatures also when $\delta\approx\Gamma$. This is apparent from
Fig.\ref{K39}, which shows the minimum temperature we measured for $^{39}K$ in
near-resonant molasses with a very low intensity of the repumping light and a large
atomic density (further details are given below). The measured temperature is well below
the Doppler limit already for $\delta<\Gamma$, although a further decrease with
increasing $\delta$ is apparent.

The second important observation is that the sub-Doppler cooling survives for detunings
larger than the Doppler cooling does, as shown for example in the calculations of
Fig.\ref{zones}. The lowest temperatures can actually be reached only for a range of
detunings where the Doppler force does no longer provide an efficient cooling. This is in
principle a problem in experiments, since the velocity capture range of the sub-Doppler
cooling mechanism is usually smaller than the initial thermal velocity, for example at
the end of the capture stage of a magneto-optical trap. Indeed, one normally needs to
exploit both Doppler and sub-Doppler cooling to achieve low temperatures \cite{josab}. We
now find that one can still combine an initial Doppler cooling with a final stage of
optimal sub-Doppler cooling by using a proper dynamical variation of detuning and
intensity of the cooling laser between the two regimes of operation. As a matter of fact,
the combination of these two ingredients makes the cooling of K as efficient as in the
other alkali species.

We now discuss in detail the experimental strategy. The linewidth of the cooling
transition for K is $\Gamma=2\pi\times$6.0 MHz, which corresponds to a Doppler
temperature $T_{D}\approx145~\mu$K. The hyperfine splitting $\Delta$ is about
3.5~$\Gamma$ and 2.2~$\Gamma$ for $^{39}$K and $^{41}$K respectively. We perform cooling
and trapping in a three-dimensional magneto-optical trap (MOT) on the $D_2$ transition
around 767~nm. The trap is loaded with pre-cooled atoms from a two-dimensional MOT. After
3~s of loading stage we have either about $2\times10^{10}$ atoms of $^{39}$K or
$4\times10^{9}$ atoms of $^{41}$K at temperatures in the $1$~mK regime. We then compress
the cloud via application of a compressed-MOT technique to densities around
$1\times10^{11}$~atoms/cm$^{3}$.
\begin{figure}[ht]
\begin{center}
\includegraphics[width=0.9\columnwidth] {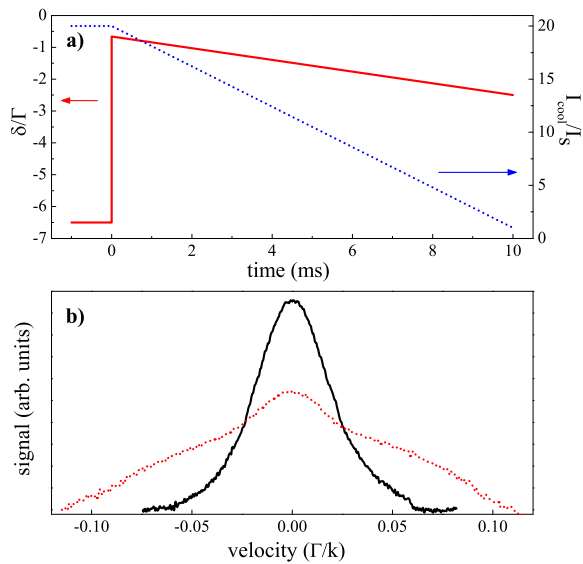}
\end{center}
\caption{Sub-Doppler cooling strategy for $^{39}$K. a) Time evolution of I$_{cool}$ and
$\delta$ (I$_{rep}$=1/100I$_{cool}$). b) Resulting velocity distribution measured after a
free expansion (black solid) compared to that obtained without the ramp (red dashed).}
\label{jump or ramp}
\end{figure}
During this initial cooling stage we adopt the standard strategy used for bosonic
potassium~\cite{fort,inouye}. We use a detuning larger than the whole excited manifold
(region IV) with total intensities as large as $20~I_{s}$ for both the cooling
(I$_{cool}$) and the repumping (I$_{rep}$) beams ($I_{s}$=1.75~mW/cm$^{2}$). This allows
for a large Doppler capture velocity. Finally, we switch off the magnetic field and we
cool the cloud in a molasses scheme as described below.

We initially reduce suddenly I$_{rep}$ to 1/100 I$_{cool}$ and we set the repumper beam
frequency on resonance with the $F=1\rightarrow F'=2$ transition. If we then try to
perform standard molasses cooling, i.e. by a sudden change of the laser parameters to the
optimal sub-Doppler cooling values, we observe moderate sub-Doppler cooling only for
small $\delta$. A larger $\delta$ results instead in a bimodal distribution of the atomic
velocities. A typical instance of such distribution is shown in Fig.~\ref{jump or ramp}b,
as measured by fluorescence imaging after a free expansion of the cloud
\cite{supplementary}. The narrow peak corresponds to sub-Doppler temperatures, while the
broader distribution can be attributed to an inefficient Doppler cooling or even to a
Doppler heating. As shown in Fig.~\ref{K39} the fraction of atoms in the central
component decreases as $\delta$ is increased. A more effective strategy consists in first
tuning the laser to $\delta\approx\Gamma/2$ to provide an initial Doppler cooling and
then in slowly decreasing the intensity while increasing $\delta$, as shown in
Fig.~\ref{jump or ramp}a. This method allows to cool nearly 90\% of the atoms to lower
temperatures, as shown in Fig.~\ref{K39}. By minimizing the final temperature we find an
optimal ramping time of about 10~ms, which corresponds to an adiabatic narrowing of the
velocity distribution during the whole sequence.
\begin{figure}[ht]
\begin{center}
\includegraphics[width=0.9\columnwidth] {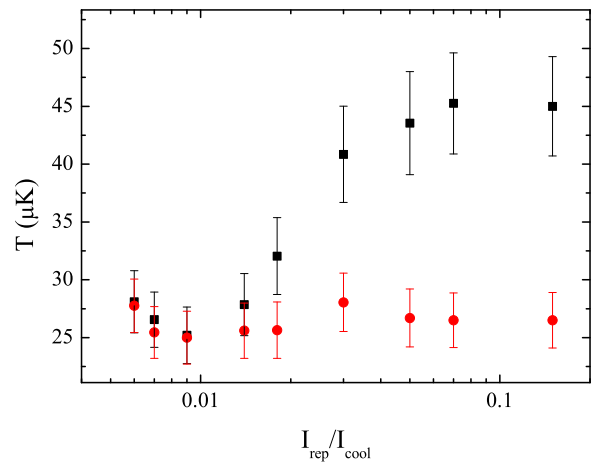}
\end{center}
\caption{Measured temperatures for $^{39}$K vs the intensity ratio of repumping and
cooling light, for densities of $4\times10^{10}$ atoms/cm$^3$ (black squares) and
$8\times10^{8}$ atoms/cm$^3$ (red dots). The heating arising from reabsorption effects at
high density can be tuned by reducing the repumper intensity.} \label{rescatt}
\end{figure}
The minimum temperature attained for $^{39}$K is about 25$\mu$K at
$\delta\approx2.5\Gamma$. It then rises again for larger $\delta$, presumably because of
the progressive reduction of the force.

We have observed, as shown in Fig.~\ref{rescatt}, that an increase of the repumping power
prevents the achievement of such low temperatures at high density, while the temperature
does not depend on the repumper power at low density. This confirms the role of
reabsorption of spontaneously emitted photons inside the cloud.

We have performed analogous measurements on $^{41}$K and found a similar behavior. In
this case a minimum temperature of about 50$\mu$K is reached for a detuning
$\delta\approx\Gamma$. We have compared the observations with a theoretical estimation of
the temperatures achievable for our experimental parameters. The optical force, shown in
Fig.~\ref{zones}, is calculated from the solution of the optical Bloch equations in the
semi-classical approximation~\cite{bambini,fort}. The model is only accurate in 1D and
for $\sigma^{+}-\sigma^{-}$ polarizations \cite{supplementary}.
\begin{figure}[ht]
\begin{center}
\includegraphics[width=0.9\columnwidth] {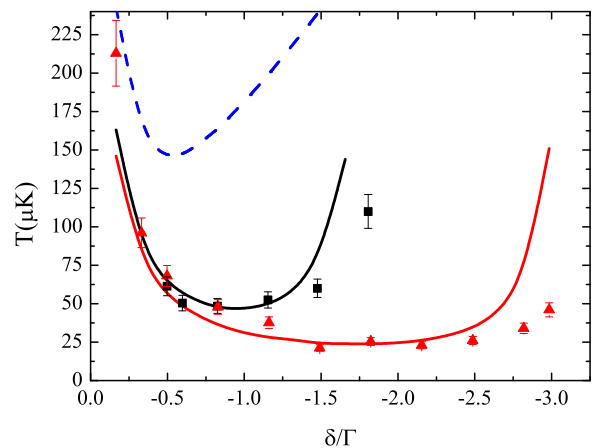}
\end{center}
\caption{Measured temperatures for $^{39}$K (red triangles) and $^{41}$K (black squares)
and calculated temperatures (lines) vs the cooling laser detuning. The dashed line is the
predicted Doppler temperature.} \label{K41}
\end{figure}
The agreement with the experimental data shown in Fig.~\ref{K41} is however rather good.
To check the validity of these calculations we directly measured the spatial diffusion
coefficient $D_x$ in the optical molasses for $^{39}$K \cite{supplementary}. By a
simultaneous measurement of the temperature we estimated the friction coefficient as
$\alpha=k_BT/D_x$~\cite{Diff}. The magnitude of $\alpha\approx 10^{-3}\hbar k^2$ is in
good agreement with the calculations. A simple estimation considering a two-level system
would give a result about 3 orders of magnitude larger. We interprete this low friction
as a result of the macroscopical occupation of the dark state.

The minimum temperatures at each detuning in Fig. \ref{K41} were obtained for an almost
constant laser intensity. The data at small detunings is apparently well described by the
scaling law observed in several other systems
\cite{scaling,castinb,erbium,strontium,ytterbium}:
\begin{equation}\label{subdoppler}
    T=C_{\sigma^+\sigma^-}\frac{\hbar\Gamma}{2k_B}\frac{\Gamma}{|\delta|}\frac{I}{I_s}+T_0.
\end{equation}
From a combined fit we get $C_{\sigma^+\sigma^-}=0.20(2)$ and $T_0=9(3)\mu$K. The
$C_{\sigma^+\sigma^-}$ coefficient is smaller than those measured on the two species
where sub-Doppler cooling has been observed for $\delta\lesssim\Gamma$, i.e. $^{87}$Er
\cite{erbium} (0.38(2)) and $^{87}$Sr \cite{strontium} (1.3(3)) but larger than those
measured in Rb and Cs at large detunings \cite{castinb}. The observed scaling for K
suggests that heating processes due to reabsorption are less efficiently suppressed when
$\delta$ is small, as expected.

To characterize the robustness of the cooling process against stray magnetic fields we
tried to keep the MOT magnetic field on during the cooling procedure. For gradients
larger than 5 G/cm, we reached the Doppler temperature. This gradient corresponds to an
average magnetic field of about 1 G, which is the same characteristic value found for the
other alkali atoms.

The techniques described here might be applied to other systems, such as the
$^1$S$\rightarrow^1$P transitions of $^{43}$Ca \cite{calcium} and $^{173}$Yb
\cite{ytterbiumb}, for which $\Delta\approx 3\Gamma$. Additionally, it would be
interesting to apply our cooling strategy to Na, for which $\Delta\approx 6 \Gamma$.
Another interesting case is that of an inverted and narrow hyperfine structure like in
$^{40}$K. In this case, there are no interfering levels which can directly cause heating.
However, an increase of $\delta$ to values of the order of $\Delta$ or more leads to a
washing out of the sub-Doppler cooling mechanism itself, since the detuning from the
various hyperfine levels becomes of the same order. We performed numerical simulations of
the atomic force and we found that this effect leads to a fast decrease of the capture
velocity with increasing detuning \cite{supplementary}. This decrease is faster than the
one for atoms with a large $\Delta$ (or with a large $\Gamma/\Delta$ as in the case of
Sr), and leads to a regime in which the minimum sub-Doppler temperature
$k_BT=D_{p}/\alpha$ exceeds the capture range. This is presumably the reason for the
bimodal distribution seen in experiments with $^{40}K$~\cite{40K}. Weaker rates of
natural depumping to dark states will possibly require forced depumping \cite{depumping}
in high density samples, but further experimental and numerical investigations are needed
\cite{simul}.

We have shown how the limitations of sub-Doppler laser cooling in atomic species with
small hyperfine splitting can be overcome by the natural control of the photon
reabsorption and adiabatic ramping of the laser parameters. The laser cooling techniques
we have developed will be easily implemented in all existing experiments with potassium
atoms. Finally, the direct application of laser cooled potassium atoms to interferometric
measurements might enable a new class of experiments ~\cite{bouyer}.

{\it Note}: we recently learned of another report of sub- Doppler temperatures in
K~\cite{note}.

We gratefully acknowledge contributions by G. Roati, S. Ferrari and F. Cataliotti. This
work is dedicated to the memory of Arturo Bambini who contributed with his deep insight
to the early investigations on ultracold potassium  in Florence. This work was supported
by INFN (MICRA collaboration), by EU (IP AQUTE), by ERC (DISQUA and QUPOL grants) and by
the ESF and CNR (EuroQUASAR program).

\section{supplementary matherial}

\section{temperature measurement}

Here we give a brief description of our detection strategy. We image the atomic cloud by
florescence imaging after time of flight. To do so we release the trap and after a
variable time we pulse the MOT beams at the maximum intensity for 30$\mu s$ (the high
intensity ensures the saturation of the atomic transition). We collect the scattered
light with an imaging system on a CCD camera. Due to the high optical density of the
sample, images taken on resonance showed saturation effects and hence prevented us from
estimating the correct number of atoms. Hence we detune the lasers on the blue of the
F=2$\rightarrow$ F'=3 transition. Detuning to the red side of the transition was avoided
due to the proximity of the other hyperfine energy levels. To determine the number of
atoms by this method we first perform spectroscopy on the atomic cloud at low density and
at the usual intensity recording the power broadening. To check the validity of this
technique we also try to perform temperature measurements tuning the laser on resonance.
Although the width of the cloud was affected by saturation effects, the resulting
temperatures were consistent with the ones measured while detuning the laser out of
resonance. We fit the atomic signal with a 2D Gaussian distribution to extract the
e$^{-1/2}$ half-width $\sigma$ of the distribution, measuring it at different expansion
times and henceforth we can reconstruct $\sigma(t_{exp})$, which we fit by the equation
\begin{equation}\label{exp}
    \sigma(t_{exp})^2=\sigma_0^2+\frac{k_B T}{m} t_{exp}^2.
\end{equation}
\begin{figure}[ht]
\begin{center}
\includegraphics[width=0.9\columnwidth] {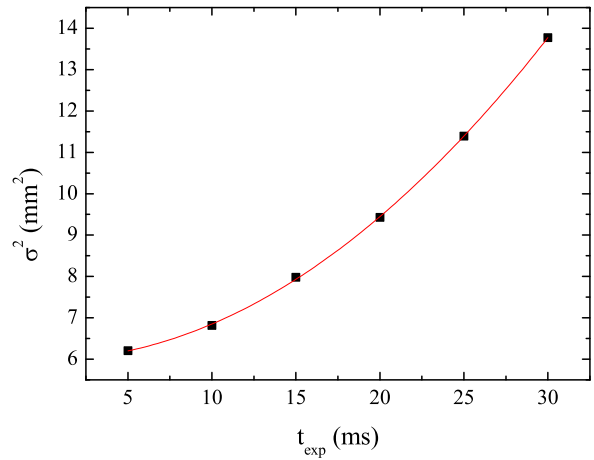}
\end{center}
\caption{Typical evolution of the atomic distribution width as a function of time after
release from the molasses cooling beams. The solid line is a fit by Eq. \ref{exp}.}
\label{sig_free}
\end{figure}
From the results of the fit we extract the temperature as well as the initial size of the
cloud, which is used in order to measure the density of the atomic cloud.

In order to precisely measure the magnification of our imaging system we use two
different methods. The first one consists of loading the atoms into a magnetic trap
mounted on a translation stage. The positioning of the translation stage has 10$\mu$m
accuracy. By movement by a given amount of the coils and recording the position of the
center of the cloud on the CCD, we are able to measure the magnification, which is found
to be: 1/3.23(1). The second method consists of allowing the atomic cloud to fall due to
gravity after switching off the trap and recording its position as a function of time.
From the fit of the position by the law of free fall and assuming for the gravitational
acceleration: g=9.81m/s$^2$ we find the value of the magnification to be 1/3.22(5), which
is in good agreement with the value determined by the other method.

In case of the bimodal distributions as in Fig. 2 of the main paper, we fit just the cold
component after a long time of flight (t$_{exp}>$10ms).
\section{diffusion measurement}
\begin{figure}[ht]
\begin{center}
\includegraphics[width=0.9\columnwidth] {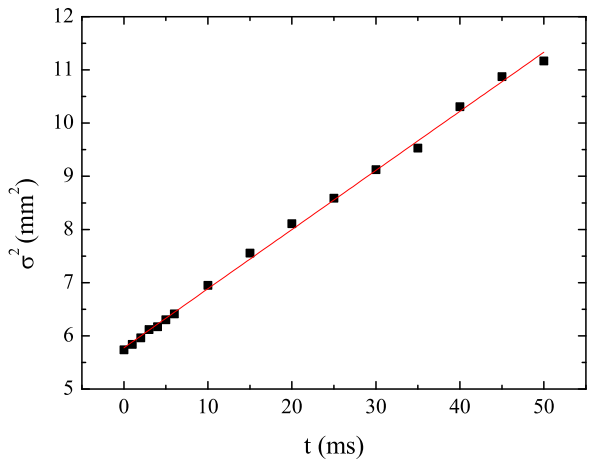}
\end{center}
\caption{Typical data for the variation of the width of the atomic distribution as a
function of the time spent inside the molasses cooling beams. The solid line is a fit by
Eq. \ref{dx}.} \label{Diff_t}
\end{figure}
We measure the spatial diffusion coefficient D$_x$ for the Brownian atomic motion during
the molasses cooling phase. This is a useful quantity since it connects to the other
quantities characterizing the cooling process (namely T, $\alpha$ and D$_p$) via the
simple formula:
\begin{equation}\label{dx}
    D_x=\frac{D_p}{\alpha^2}=\frac{k_B T}{\alpha}.
\end{equation}
\begin{figure}[ht]
\begin{center}
\includegraphics[width=0.9\columnwidth] {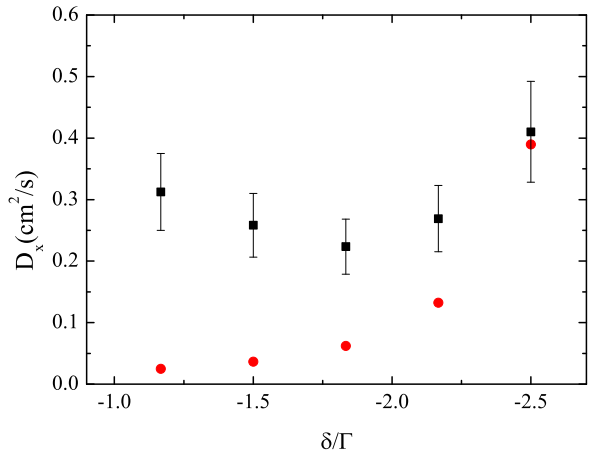}
\end{center}
\caption{Measured diffusion coefficients in the optical molasses (black squares) compared
to the numerically calculated values (red dots) as a function of the detuning of the
cooling laser. } \label{Diff}
\end{figure}
Hence, by measuring both T and D$_x$ one can reconstruct the other useful quantities. The
measurement is performed at different final detuning of the molasses cooling beams by
allowing the atomic cloud to expand in presence of the cooling light for a variable time,
and then taking an image after 100$\mu$s dark period. By recording the variation of the
width with time in the molasses cooling beams, we find it to be in agreement with a
diffusion process (Fig. \ref{Diff_t}). By fitting the evolution of the cloud size with
the equation:
\begin{equation}\label{brown}
    \sigma(t)^2=\sigma_0^2+2D_xt
\end{equation}
we are able to extract the spatial diffusion coefficient. From the results in Fig.
\ref{Diff} we see that the agreement with the theory is good for our usual experimental
parameters but the agreement becomes rather poor for smaller detunings. This might be due
to the presence of additional heating from rescattered photons since these measurements
were taken at high densities (4$\times$10$^{10}$ atoms/cm$^3$). These values for the
diffusion coefficient are a factor about of 1000 times higher than the ones measured in
\cite{Diff}. The reason for this is the very low population in the F=2 level caused by
natural depumping and the use of a very weak repumping light.
\section{Numerical simulation}
We perform numerical simulations of the optical Bloch equations for multilevel systems to
find the optical forces acting on the atoms during the molasses cooling. We work in 1D
and in the $\sigma_{+}-\sigma_{-}$ configuration for the laser beams polarizations. This
is only an approximation to the more complex polarization geometries arising in the 3D
laser configuration we have in our experiment. In practical situations, in a
magneto-optical trap, both Sisyphus cooling (lin$\perp$lin) and $\sigma_{+}-\sigma_{-}$
polarization gradient cooling play a role in the cooling process, with the former
dominating at large detuning and the latter at small detuning \cite{josab, pol}. In our
simulation, only the cooling beam is taken into account to find the cooling force. This
is justified by the very low repumper intensity we use in the experiment (about
10$^{-2}$I$_{s}$).

In the determination of the different regions of Fig. 1 of the main paper we focus our
attention on a velocity range corresponding to twice the width of our initial velocity
distribution (just before the molasses phase the atomic temperature is about 1 mK) and we
set the cooling laser intensity to be I$_s$. The cooling force was calculated as a
function of velocity for different detunings. The whole range of detunings for which the
force was calculated was divided into four regions. Region I: This region comprises of
the detuning values for which the force is opposite to velocity for the whole velocity
range and hence always provides cooling. Region II: The force is opposite to the velocity
only for small velocities, whereas for higher velocities the force changes sign, thereby
providing heating. Region III: In this region, the force provides heating in the whole
velocity range. Region IV: The force provides heating for low velocities, whereas for
higher velocities the force is opposite to the velocity. In Fig. 1 of the main paper we
do not report the force behavior for detuning values in between the F'=2 and F'=0 levels.

The calculated temperatures of Fig.5 in the main paper are calculated as
$k_BT=D_{p}/\alpha$ where $D_{p}$ is the momentum diffusion coefficient and $\alpha$ is
the friction coefficient calculated as the slope of the force at $v=0$. We estimate $D_p$
by using the simple argument for the random step of the Brownian motion in momentum space
in multilevel transitions described in Ref.~\cite{castin}.
\begin{figure}[ht]
\begin{center}
\includegraphics[width=0.9\columnwidth] {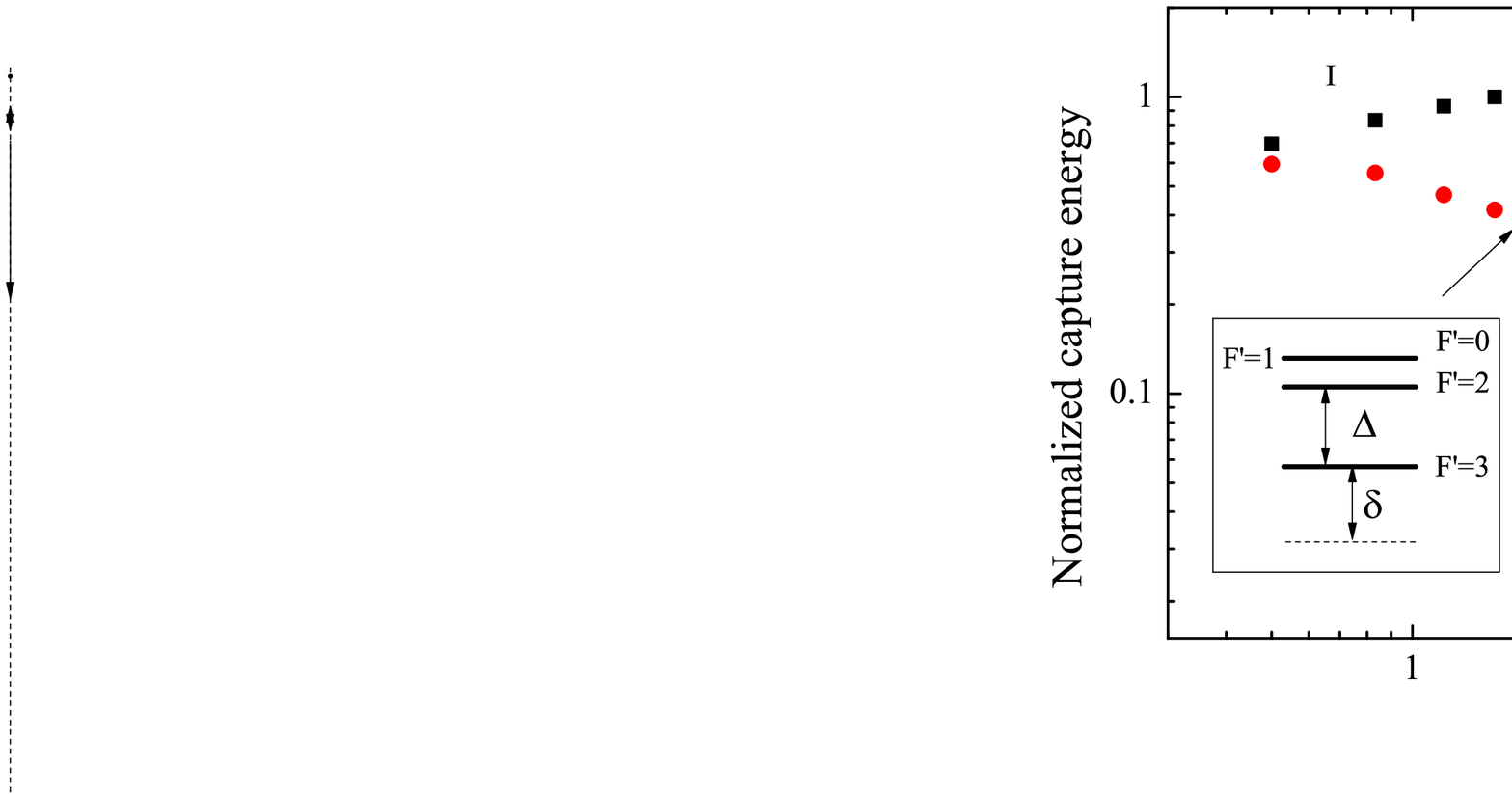}
\end{center}
\caption{Calculated sub-Doppler capture energy for $^{39}$K (black squares) and for a
hypothetical atomic species with an inverted hyperfine structure as shown in the inset
(red dots). Both quantities are normalized to the capture energy of an atomic species
with $\Delta$=35$\Gamma$.} \label{Tc}
\end{figure}

To compare the efficiency of the cooling process in the case of narrow hyperfine
structure to the two-level case we numerically simulate the cooling force and extract the
velocity capture range v$_c$, defined as the velocity giving the first local maximum of
force at low velocity. We do this in three cases: a) a hyperfine level structure like the
one of $^{39}$K, b) an equal but inverted hyperfine level structure, c) the same
hyperfine structure but with a 10-fold increase in the hyperfine splitting. In
Fig.\ref{Tc} we plot the capture energy $E_c=\frac{1}{2}m v_c^2$ for the case of $^{39}$K
normalized by the 10-fold increased case (black squares), and the inverted case
normalized by the same quantity (red dots). We show only the value of E$_C$ in the
detuning regions in which the sub-Doppler cooling mechanism provides cooling (region I
and II of Fig.1 in the main paper). This fast decrease of the velocity capture range
makes the achievement of sub-Doppler cooling difficult for optical molasses at large
detuning for this kind of systems. In conclusion, the numerical simulation provides a
good understanding of the sub-Doppler cooling process and the results obtained in our
experiment.

\end{document}